\documentclass[final]{elsarticle}

\usepackage{graphicx}
\usepackage{color}
\usepackage{dcolumn}
\usepackage{bm}

\begin{document}

\title{High-precision measurement of the proton elastic form factor ratio $\mu_{p} G_E/G_M$ at low $Q^2$}
\author[mit,anl]{X.~Zhan}
\author[uk]{K.~Allada}
\author[wm]{D.~S.~Armstrong}
\author[anl]{J.~Arrington}
\author[mit]{W.~Bertozzi}
\author[fiu]{W.~Boeglin}
\author[jlab]{J.-P.~Chen}
\author[uva]{K.~Chirapatpimol}
\author[seoul]{S.~Choi}
\author[jlab]{E.~Chudakov}
\author[infn1,infn2]{E.~Cisbani}
\author[smith]{P.~Decowski}
\author[uk]{C.~Dutta}
\author[infn1]{S.~Frullani}
\author[ubp]{E.~Fuchey}
\author[infn1]{F.~Garibaldi}
\author[mit]{S.~Gilad}
\author[jlab,rutgers]{R.~Gilman}
\author[stmary,dalhousie]{J.~Glister}
\author[anl]{K.~Hafidi}
\author[wm]{B.~Hahn}
\author[jlab]{J.-O.~Hansen}
\author[jlab]{D.~W.~Higinbotham}
\author[longwood]{T.~Holmstrom}
\author[anl]{R.~J.~Holt}
\author[mit]{J.~Huang}
\author[regina]{G.~M.~Huber}
\author[ubp]{F.~Itard}
\author[jlab]{C.~W.~de Jager}
\author[rutgers]{X.~Jiang}
\author[nwu]{M.~Johnson}
\author[wm]{J.~Katich}
\author[infn3]{R.~de~Leo}
\author[jlab]{J.~J.~LeRose}
\author[uva]{R.~Lindgren}
\author[kent]{E.~Long}
\author[csla]{D.~J.~Margaziotis}
\author[nrcn]{S.~May-Tal~Beck}
\author[jlab]{D.~Meekins}
\author[jlab]{R.~Michaels}
\author[mit,jlab]{B.~Moffit}
\author[uva]{B.~E.~Norum}
\author[stnorbert]{M.~Olson}
\author[telaviv]{E.~Piasetzky}
\author[telaviv]{I.~Pomerantz}
\author[glasgow]{D.~Protopopescu}
\author[duke]{X.~Qian}
\author[duke,jlab]{Y.~Qiang}
\author[syracuse]{A.~Rakhman}
\author[rutgers]{R.~D.~Ransome}
\author[anl]{P.~E.~Reimer}
\author[fiu]{J.~Reinhold}
\author[uva]{S.~Riordan}
\author[telaviv,lbnl]{G.~Ron}
\author[jlab]{A.~Saha}
\author[stmary]{A.~J.~Sarty}
\author[jlab,temple]{B.~Sawatzky}
\author[rutgers]{E.~C.~Schulte}
\author[uva]{M.~Shabestari}
\author[yerevan]{A.~Shahinyan}
\author[telaviv]{R.~Shneor}
\author[ijs,ul]{S.~\v{S}irca}
\author[anl,jlab]{P.~Solvignon}
\author[mit,temple]{N.~F.~Sparveris}
\author[usc]{S.~Strauch}
\author[uva]{R.~Subedi}
\author[mit,jlab]{V.~Sulkosky}
\author[infn3]{I.~Vilardi}
\author[ill]{Y.~Wang}
\author[jlab]{B.~Wojtsekhowski}
\author[hampton]{Z.~Ye}
\author[lanzhou]{Y.~Zhang}

\address[mit]{Massachusetts Institute of Technology, Cambridge, MA 02139, USA}
\address[anl]{Physics Division, Argonne National Laboratory, Argonne, IL 60439, USA}
\address[uk]{University of Kentucky, Lexington, KY 40506, USA}
\address[wm]{College of William and Mary, Williamsburg, VA 23187, USA}
\address[fiu]{Florida International University, Miami, FL 33199, USA}
\address[jlab]{Thomas Jefferson National Accelerator Facility, Newport News, VA 23606, USA}
\address[uva]{University of Virginia, Charlottesville, VA 22904, USA}
\address[seoul]{Seoul National University, Seoul 151-747, Korea}
\address[infn1]{INFN, Sezione di Roma - Gruppo Sanit\`{a}, I-00161 Rome, Italy}
\address[infn2]{Istituto Superiore di Sanit\`{a}, I-00161 Rome, Italy}
\address[smith]{Smith College, Northampton, MA 01063, USA}
\address[ubp]{Universit\'{e} Blaise Pascal, F-63177 Aubiere, France}
\address[rutgers]{Rutgers, The State University of New Jersey, Piscataway, NJ 08855, USA}
\address[stmary]{Saint Mary's University, Halifax, Nova Scotia, B3H 3C3, Canada}
\address[dalhousie]{Dalhousie University, Halifax, Nova Scotia, B3H 3J5, Canada}
\address[longwood]{Longwood University, Farmville, VA 23909, USA}
\address[regina]{University of Regina, Regina, SK, S4S 0A2, Canada}
\address[nwu]{Northwestern University, Evanston, IL 60208, USA}
\address[infn3]{Dipartimento di Fisica and INFN sez. Bari, Bari, Italy}
\address[kent]{Kent State University, Kent State, OH 44242, USA}
\address[csla]{California State University Los Angeles, Los Angeles, CA 90032, USA}
\address[nrcn]{NRCN, P.O.Box 9001, Beer-Sheva, 84190, Israel}
\address[stnorbert]{Saint Norbert College, De Pere, WI 54115, USA}
\address[telaviv]{Tel Aviv University, Tel Aviv 69978, Israel}
\address[glasgow]{University of Glasgow, Glasgow G12 8QQ, Scotland, UK}
\address[duke]{Duke University, Durham, NC 27708, USA}
\address[syracuse]{Syracuse University, Syracuse, NY 13244, USA}
\address[lbnl]{Lawrence Berkeley National Lab, Berkeley, CA 94720, USA}
\address[temple]{Temple University, Philadelphia, PA 19122, USA}
\address[yerevan]{Yerevan Physics Institute, Yerevan 375036, Armenia}
\address[ijs]{Institute ``Jo\v{z}ef Stefan'', 1000 Ljubljana, Slovenia}
\address[ul]{Dept. of Physics, University of Ljubljana, 1000 Ljubljana, Slovenia}
\address[usc]{University of South Carolina, Columbia, SC 29208, USA}
\address[ill]{University of Illinois at Urbana-Champaign, Urbana, IL 61801, USA}
\address[hampton]{Hampton University, Hampton, VA 23668, USA}
\address[lanzhou]{Lanzhou University, Lanzhou, China}

\date{\today}

\begin{abstract}

We report a new, high-precision measurement of the proton elastic form factor
ratio $\mu_pG_E/G_M$ for the four-momentum transfer squared $Q^2$ = 0.3--0.7~
(GeV/$c$)$^2$. The measurement was performed at Jefferson Lab (JLab) in Hall A
using recoil polarimetry. With a total uncertainty of approximately 1\%, the
new data clearly show that the deviation of the ratio $\mu_pG_E/G_M$ from
unity observed in previous polarization measurements at high $Q^2$ continues
down to the lowest $Q^2$ value of this measurement. The updated global fit
that includes the new results yields an electric (magnetic) form factor
roughly 2\% smaller (1\% larger) than the previous global fit in this
$Q^2$ range. We obtain new extractions of the proton electric and magnetic
radii, which are $\langle r^2_{E} \rangle^{1/2}=0.875\pm0.010$ fm and $\langle
r^2_{M} \rangle^{1/2}=0.867\pm0.020$ fm. The charge radius is consistent with
other recent extractions based on the electron-proton interaction, including
the atomic hydrogen Lamb shift measurements, which suggests a missing
correction in the comparison of measurements of the proton charge radius using
electron probes and the recent extraction from the muonic hydrogen Lamb shift.

\end{abstract}


\maketitle

The nucleon electromagnetic form factors are fundamental quantities which
relate to the charge and magnetization distributions within the nucleon. For
over 40 years, the form factors have been studied extensively by Rosenbluth
separations of the unpolarized electron-proton
scattering cross section. Recent advances in the technology of
intense polarized beams, polarized targets and polarimetry ushered in a
new generation of experiments that measure double polarization
observables~\cite{arrington07a,perdrisat07,arrington11a}. Although the proton
electric to magnetic form factor ratio $R \equiv \mu_pG_E/G_M$ determined by
unpolarized measurements showed minimal $Q^2$ dependence up to $Q^2\approx
6$~(GeV/$c$)$^2$, experiments at JLab with high-quality polarized electron
beams measuring recoil polarization~\cite{gayou02, punjabi05,
puckett10} revealed that the ratio $\mu_pG_E/G_M$ drops almost
linearly with $Q^2$ for $Q^2$$>$$1$~(GeV/$c$)$^2$. These findings
led to an explosion of experimental and theoretical efforts to understand the
proton form factors~\cite{qattan05, perdrisat07, arrington11a}. The difference
between the polarization and cross section measurements is now believed to be
the result of two-photon exchange (TPE) contributions~\cite{guichon03,
blunden03, chen04, arrington11b}, which have little impact on the polarization
measurements but significantly affect the Rosenbluth extractions of $G_E$ at
large $Q^2$.

While measurements at large momentum transfer have provided information on
the fine details of the proton structure and relate to the quark orbital
angular momentum, the low $Q^2$ form factor behavior is sensitive to the
long-range structure which is believed to be dominated by the ``pion cloud''.
High-precision measurements at low $Q^2$ were motivated by a recent
semi-phenomenological fit~\cite{friedrich03}, which suggested that structure
might be present in all four nucleon form factors at $Q^2\approx
0.3$~(GeV/$c$)$^2$. Later polarization measurements from
MIT-Bates~\cite{crawford07} and JLab~\cite{ron07} probed this region with
a precision of $\sim$2\%, but found no indication of such structure in the
ratio $\mu_pG_E/G_M$.

This Letter reports on a new, high-precision polarization transfer measurement
(JLab E08-007) of the proton form factor ratio $\mu_pG_E/G_M$ at $Q^2$ values
between 0.3 and 0.7~(GeV/$c$)$^2$ . In the one-photon exchange (Born)
formalism, the ratio of the transferred transverse to longitudinal
polarizations is related to the proton form factors~\cite{perdrisat07}:
\begin{equation}
R\equiv \mu_p\frac{G_E}{G_M}=-\mu_p\frac{E_e+E_e'}{2M_p}\tan(\frac{\theta_e}{2})\frac{P_t}{P_l},
\label{eq:form}
\end{equation}
where $M_p$ and $\mu_p$ are the proton mass and magnetic moment, $E_e$
$(E_e')$ is the incident (scattered) electron energy, $\theta_e$ is the
electron scattering angle and $P_t$ ($P_l$) is the transverse (longitudinal)
component of the polarization transfer.

The experiment was performed at Jefferson Lab in Hall A~\cite{alcorn04}. A
1.2 GeV polarized electron beam of between 4 and 15 $\mu$A was
incident on a 6 cm thick liquid hydrogen target. The beam helicity was changed
at 30 Hz, with a quartet structure selected pseudorandomly between ($+--+$) and
($-++-$) for each set of four helicity states.  The beam polarization was
near 83\%, as measured by the M{\o}ller polarimeter in the Hall~\cite{alcorn04}.
The recoil proton was detected by the left High Resolution Spectrometer (LHRS)
in coincidence with the elastically scattered electron detected in a large
acceptance spectrometer (``BigBite''). The transferred proton polarization was
measured by a focal plane polarimeter (FPP) in the LHRS~\cite{alcorn04}.
The trigger was a coincidence between detection of a charged particle in the
HRS and a signal in the BigBite calorimeter used to select energetic electrons.
Since elastic events can be well identified using the proton kinematics, only
information from the BigBite calorimeter was used in the analysis;
the tracking detectors were not turned on for the experiment. The kinematic
settings are given in Table~\ref{tab:kin}.

\begin{table}
\caption{Kinematic settings for the experiment. $\theta_e~(\theta_p)$ is
the scattered electron (proton) angle in the lab frame, $P_p$ is the proton
central momentum, $\varepsilon$ is the virtual photon polarization. }
\label{tab:kin}
\begin{tabular}{ccccccc}
$Q^2$ & $\theta_e$& $\theta_p$& $P_p$& $\varepsilon$ & Analyzer\\
(GeV/$c$)$^2$ & (deg) & (deg) & (GeV/$c$) & & thickness\\
\hline
0.298 & 28.5 & 60.0 & 0.565 & 0.88 & 2.25"\\
0.346 & 31.3 & 57.5 & 0.616 & 0.85 & 2.25"\\
0.402 & 34.2 & 55.0 & 0.668 & 0.82 & 3.75"\\
0.449 & 36.7 & 53.0 & 0.710 & 0.80 & 3.75"\\
0.494 & 39.2 & 51.0 & 0.752 & 0.78 & 3.75"\\
0.547 & 41.9 & 49.0 & 0.794 & 0.75 & 3.75"\\
0.599 & 44.6 & 47.0 & 0.836 & 0.72 & 3.75"\\
0.695 & 49.8 & 43.5 & 0.913 & 0.66 & 3.75"\\
\end{tabular}
\end{table}

The FPP measured the azimuthal asymmetry due to the spin-orbit coupling in
proton scattering from carbon nuclei. The proton's transferred polarization
was extracted by a weighted-sum method~\cite{besset79a} based on a
COSY~\cite{cosy} model of the spin precession in the spectrometer. Due to the
fast reversal of beam helicity, the helicity-independent corrections
(acceptance, detector efficiency, target density, etc.) cancel. At each $Q^2$
value, measurements were taken at the nominal elastic kinematics and with the
HRS spectrometer momentum shifted by +2\% and -2\% to allow for additional
systematic studies of the spin transport model. The kinematic factors in
Eq.~\ref{eq:form} were determined from the beam energy $E_e$ and the proton
scattering angle $\theta_p$.

A series of cuts were applied to cleanly select the elastic events and 
minimize the systematic uncertainties. In addition to the standard HRS cuts
(tracking, acceptance, timing, etc.), cuts on the FPP variables were applied
to select the events with reliable second-scattering
reconstruction~\cite{xiaohui_thesis}. For elastic event selection, we use the
correlation between the proton angle and momentum. After applying a tight cut
(1.7$\sigma$) on the elastic peak and a target vertex cut, the contamination
from the aluminum endcaps of the target and neutral pion photoproduction is
less than 0.1\% in the extracted ratio $\mu_pG_E/G_M$, which is negligible
compared to other systematic uncertainties.

The primary systematic uncertainty in the measurement comes from uncertainty
in the spin precession of the protons in the spectrometer magnetic fields. The
uncertainty arising from imperfect knowledge of the spectrometer fields was
estimated by examining the change in the results when key parameters of
the spectrometer optics model used for spin precession were altered. The most 
sensitive parameter is the dipole central bending angle $\Theta_0$, which
is nearly 45$^\circ$. A
conservative 5.5 mrad uncertainty of this parameter is quoted from previous
analysis~\cite{punjabi05}, yielding a 0.1--0.6\% change in
the ratio for different kinematics. In addition, any error in the particle
trajectory reconstruction will lead to incorrect proton kinematics at the
target which are used as input to the spin precession calculation. The largest
effect comes from offsets in the proton angle in the scattering plane,
estimated to be $\sim$1~mrad based on the consistency of the particle
kinematics with the elastic constraint when accounting for the uncertainties
of other related parameters (beam energy, proton momentum, spectrometer angle).
The associated uncertainty on the extracted form factor ratio
is determined to be 0.6--0.9\% by manually shifting the proton trajectories in
the spin transport~\cite{xiaohui_thesis}.  As a final check, the extracted
value for $\mu_pG_E/G_M$ was examined as a function of each of the
reconstructed proton target variables and for each of the three momentum
settings taken for each $Q^2$ setting.  All of these results were consistent
within the statistics of the measurements.

\begin{table}
\caption{\label{tab:results} Experimental ratio $\mu_pG_E/G_M\pm
stat.\pm syst.$ along with the average FPP analyzing power $\langle A_c
\rangle$, efficiency $\epsilon$ and the figure of merit ($\epsilon A_c^2$)
for secondary scattering angles between $5^{\circ}$ and $25^{\circ}$.}
\begin{tabular}{ccccc}
$Q^2$		& $\langle A_c \rangle$	& $\epsilon$	& FOM	& $\mu_pG_E/G_M$ \\
(GeV/$c$)$^2$ 	&  			& [\%]		& [\%]	&  \\
\hline
       0.298  & 0.219 & 5.30 & 0.25 & $0.9272$$\pm$$0.0114$$\pm$$0.0071$\\
       0.346  & 0.394 & 3.67 & 0.57 & $0.9433$$\pm$$0.0088$$\pm$$0.0093$\\
       0.402  & 0.466 & 4.36 & 0.95 & $0.9318$$\pm$$0.0066$$\pm$$0.0076$\\
       0.449  & 0.488 & 4.09 & 0.97 & $0.9314$$\pm$$0.0060$$\pm$$0.0073$\\
       0.494  & 0.466 & 3.81 & 0.83 & $0.9286$$\pm$$0.0054$$\pm$$0.0076$\\
       0.547  & 0.430 & 4.34 & 0.81 & $0.9274$$\pm$$0.0055$$\pm$$0.0071$\\
       0.599  & 0.392 & 4.41 & 0.68 & $0.9084$$\pm$$0.0053$$\pm$$0.0104$\\
       0.695  & 0.334 & 4.74 & 0.53 & $0.9122$$\pm$$0.0045$$\pm$$0.0107$\\
\end{tabular}
\end{table}

The experimental results are summarized in Table~\ref{tab:results}. The
average FPP analyzing power $\langle A_c \rangle$ and efficiency $\epsilon$
are in good agreement with previous measurements~\cite{glister09}.  When
a transversely polarized proton interacts with the FPP, the azimuthal
asymmetry induced by the secondary scattering is proportional to the proton
polarization and the analyzing power, which depends on the proton momentum
and the angle of the
secondary scattering.  The average analyzing power extracted in the analysis
is the event-weighted average over the 5--25$^\circ$ range of angles used
in the analysis of the experiment.  The FPP figure of merit (FOM) is an
integral of $\epsilon A_c^2$ over the selected $\theta$ range. No correction
was applied for two-photon exchange, which is less than a 0.1\% correction on
$P_t/P_l$ (or $G_E/G_M$) for these kinematics~\cite{blunden03}.

With the unprecedented statistical precision, we were able observe a small cut
dependence in the analysis when loose cuts (larger than 2$\sigma$) were placed
on the elastic peak, and so used tighter cuts (1.7$\sigma$) than previous
experiments where were statistics limited.  This also led to a detailed
reanalysis of a previous experiment~\cite{ron07}, which found that their
sensitivity to these effects were small but yielded an improved isolation and
correction for events involving scattering from the cryotarget endcaps that
led to a slight decrease in the extracted form factor ratio~\cite{ron11}.

\begin{figure}[ht]
\begin{center}
\includegraphics[angle=0,width=.6\textwidth]{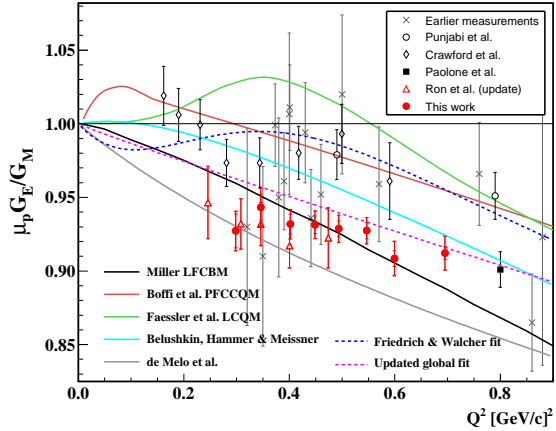}
\caption[]{\label{fig:ratio} Low $Q^2$ polarization measurements of
$\mu_pG_E/G_M$. The solid red circles are results from E08-007
and the hollow symbols are from previous measurements~\cite{punjabi05,
crawford07, ron11, paolone10} with precision of better than 3\%.
The grey crosses show various lower precision data~\cite{milbrath98,
pospischil01, dieterich01, gayou01}. The solid curves are selected theoretical
models of the proton form factors~\cite{miller02a, boffi02, faessler06,
belushkin07, melo09}. The dashed curves are the Friedrich and Walcher
fit~\cite{friedrich03}, and the updated global fit (see text).}
\end{center}
\end{figure}

Figure~\ref{fig:ratio} presents the new results together with previous
polarization measurements, including the updated results from Ron, \textit{et
al}~\cite{ron11}.  The inner error bars are the statistical uncertainty and
the outer error bars are the statistical and systematic uncertainties added in
quadrature. The new results achieved a total uncertainty of $\sim$1\%, usually
dominated by the systematic uncertainties. The data fall slowly across the
covered $Q^2$ range and strongly deviate from unity. The high-precision
point at $Q^2=0.8$~(GeV/$c$)$^2$~\cite{paolone10} is also consistent with the
trend of our new results. Compared to the Bates measurement~\cite{crawford07},
these new ratio results are lower by 2--3$\sigma$, indicating some systematic
discrepancy.

Two implications of the new data can be directly seen.  Some previous
measurements suggested that the ratio was flat at $\mu_pG_E/G_M
\approx 1$ until $Q^2 \approx 0.2$~(GeV/$c$)$^2$ and then began to fall, which
implied identical low-$Q^2$ behavior of the form factors and thus equal
charge and magnetization radii for the proton.  We observe that $\mu_pG_E/G_M$
is significantly below one over the entire $Q^2$ range of our data, with no
indication that the ratio is approaching unity noticeably above $Q^2=0$. The
global fit (described later) suggests that the ratio begins to fall starting
at or very near $Q^2=0$, implying significant differences in the large scale
spatial distribution of charge and magnetization in the proton.  Second, with
this high precision data, we see no indication of any structure in the ratio
$\mu_pG_E/G_M$ over the covered $Q^2$ region.

The curves in Fig.~\ref{fig:ratio} show a selection of modern fits and models
in the literature: a light-front cloudy-bag model (Miller~\cite{miller02a}), a
point-form chiral constituent quark model (Boffi {\it et
al.}~\cite{boffi02}), a Lorentz covariant constituent quark model (Faessler
{\it et al.}~\cite{faessler06}), and two representative vector-meson dominance
models (Belushkin {\it et al.}~\cite{belushkin07} and de Melo {\it et
al.}~\cite{melo09}). None of the existing models precisely match the new
results, but we note that despite utilizing different theoretical frameworks,
both the calculation of Miller and of de Melo {\it et al.} are in
qualitatively good agreement with our results.  As both of these consider
the role of non-valence contributions, e.g. nucleon plus pion contributions
in addition to the lowest 3-quark Fock state, this provides additional
support for the important role of the pion cloud at low $Q^2$, even though
no structure is seen in this ratio (or in the global fit for $G_E$ and $G_M$).

\begin{figure}[ht]
\begin{center}
\includegraphics[angle=0,width=.55\textwidth]{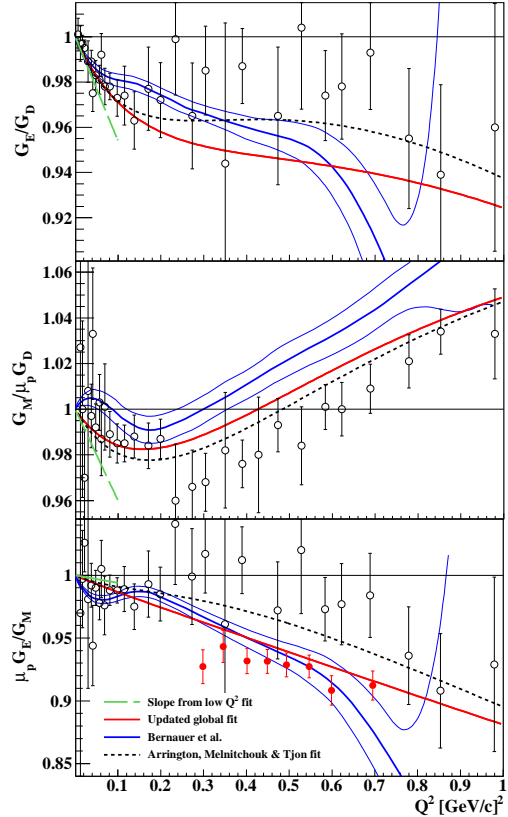}
\caption[]{\label{fig:ff} Global fits to $G_E$, $G_M$ and ratio
$\mu_pG_E/G_M$ from ref.~\cite{arrington07c} (dotted line) and this work
(solid red line). The blue lines are the best fit and uncertainty band from
the Mainz cross section data~\cite{bernauer10}. The solid circles are our new
results for $\mu_pG_E/G_M$ and the hollow circles show the global cross section
analysis of Ref.~\cite{arrington07c}, which do \textit{not} include TPE TPE
corrections.  The green long-dashed line shows the first term of our low-$Q^2$
fit used to extract the radius.}
\end{center}
\end{figure}

We have performed an updated global fit of the form factors, following the
procedure of Ref.~\cite{arrington07c}.  In addition to the data sets
included in that analysis, we add our new results along with other recent
polarization measurements~\cite{puckett10, ron11, paolone10}, but have not
included the more recent cross section measurements from
Mainz~\cite{bernauer10}, as these data do not include sufficient information
on the uncertainties to be included in a consistent fashion.
Figure~\ref{fig:ff} presents $G_E$ and $G_M$ normalized by the standard dipole
form, $G_D = 1/[1+Q^2/0.71(\mathrm{GeV}/c)^2]^2$, for the previous global
fit and our updated fit, along with the Rosenbluth separation value for the
cross section analysis of Ref.~\cite{arrington07c} without the inclusion of
TPE corrections.  The difference between the cross section
analysis (hollow circles) and the global fit (dotted curve), both from
Ref.~\cite{arrington07c}, is due to a combination of the TPE corrections and
the inclusion of the limited polarization data available at the time.
The difference between the dotted curve and our updated fit (solid red curve)
comes from the inclusion of the new polarization data (Refs.~\cite{paolone10,
ron11} and this work). For 0.2$<$$Q^2$$<$1.0~(GeV/$c$)$^2$, the updated global
fit shows a 2\% decrease in $G_E$ and a 1\% increase in $G_M$ compared to the
earlier fit~\cite{arrington07c}. The inclusion of these precise polarization
data also improves the determination of the relative normalization of the
different cross section data sets included in the fit, which has a significant
impact on the uncertainty in the extraction of the proton charge
radius~\cite{sick03}.

We also show the fit to recent Mainz cross section
measurements~\cite{bernauer10} in Fig.~\ref{fig:ff}. The Mainz fit is in
reasonable agreement with our direct extraction of $\mu_pG_E/G_M$, but their
results for $G_E$ and especially $G_M$ are somewhat inconsistent with our
global fit (and systematically above all previous extractions of $G_M$). 
However, it is difficult to make a detailed comparison of the results due to
differences in TPE corrections.  The hollow points do not include any TPE
correction, the global fits include the full TPE corrections of
Ref.~\cite{blunden03} to the cross sections, while the Mainz
extraction~\cite{bernauer10} applies only the $Q^2=0$ limit of the Coulomb
distortion correction~\cite{mckinley48}. This yields an overestimate of the
correction at all finite $Q^2$ values and neglects the $Q^2$ dependence of the
correction which will change the extracted magnetic form factor and magnetic
radius~\cite{arrington11c,bernauer11,ron11}, but has a much smaller impact
on the very low $Q^2$ behavior of $G_E$ and the extracted charge radius. A full
examination of the consistency between the Mainz result and other cross
section measurements would require careful examination of the TPE corrections
applied to the different data sets and is beyond the scope of the present work.

The proton electric and magnetic RMS radii can be determined from the form
factor slope at $Q^2=0$:
\begin{equation}
\langle r^2_{E/M} \rangle = - \frac{6}{G_{E/M}(0)}\left(\frac{dG_{E/M}(Q^2)}{dQ^2}\right)_{Q^2=0}.
\end{equation}
We do not use the previously described global fit to extract the radius
because such fits are dominated by high-$Q^2$ data which have essentially no
sensitivity to the radius.  To extract the radius, we repeat the global fit
described above using a continued fraction~\cite{sick03} parameterization with
fewer parameters, including only cross section and polarization data below
$Q^2$=0.5~(GeV/$c$)$^2$. The radius values we obtain are:
\begin{eqnarray}
\langle r^2_{E} \rangle^{1/2} &=& 0.875\pm0.008_{\mathrm{exp}}\pm0.006_{\mathrm{fit}}~\mathrm{fm}\\
\langle r^2_{M} \rangle^{1/2} &=& 0.867\pm0.009_{\mathrm{exp}}\pm0.018_{\mathrm{fit}}~\mathrm{fm},
\end{eqnarray}
where the quoted values are the averages from fits with 3, 4, and 5
parameters, the first uncertainty is the result of the experimental
uncertainties (statistical, systematic and normalization uncertainties added
in quadrature) and the second is the model dependence estimated by taking the
scatter of fits when we vary the functional form, number of fitting
parameters, and $Q^2$ cutoff. The slope corresponding to first term of the
low-$Q^2$ analysis is shown in Fig.~\ref{fig:ff}.  For $G_E$, the global fit
and low-$Q^2$ fit are in good agreement as $Q^2 \to 0$, although by $Q^2=0.1$,
the higher-order terms are beginning to yield a deviation from the linear
term. For $G_M$, even the slope at $Q^2=0$ is slightly different. Because
the cross section becomes more sensitive to $G_M$ as $Q^2$ increases,
fits that include high $Q^2$ data will tend to overemphasize agreement at
high $Q^2$ values at the cost of the low $Q^2$ behavior.  This is why the
global fits are generally not well suited to extraction of the radii,
especially when fitting the magnetic form factor.

While the charge and magnetic radii yield consistent values, as one would
expect in the non-relativistic limit where the quarks carry both the
charge and magnetization, the fit yields a different detailed behavior of the
form factors even at low $Q^2$.  This implies a different shape for the charge
and magnetization distributions, even though the radii are consistent.  If a
fixed functional form is chosen for both the charge and magnetic form factors
at low $Q^2$, the fits yield different radii.

Our magnetic radius is significantly larger than the Mainz
value~\cite{bernauer10}: $\langle r^2_M \rangle^{1/2} = 0.777\pm0.017$~fm. As
noted previously, the Coulomb correction applied in that analysis does not
include any $Q^2$ dependence, which will likely lead to an error in the
extracted magnetic radius.  An analysis of their data with a modern correction
for Coulomb distortion (or full two-photon exchange) is required to determine
if this represents a true discrepancy between the Mainz result and all other
extractions.  An initial evaluation~\cite{arrington11c,bernauer11}, using
a calculation for the TPE corrections valid up to $Q^2 \approx 1.0$~GeV$^2$,
yields an increase in the magnetic radius of 0.026~fm, accounting for
roughly one-third of the discrepancy.

\begin{figure}[ht]
\begin{center}
\includegraphics[angle=0,width=.6\textwidth]{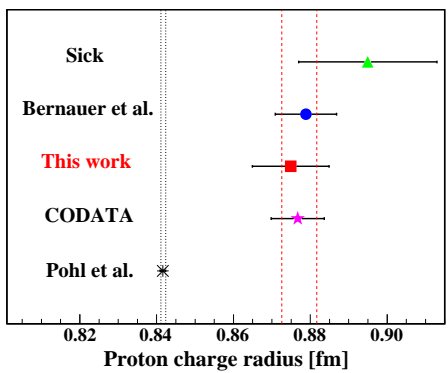}
\caption[]{\label{fig:radius} The proton RMS charge radius from
a previous $ep$ scattering analysis (Sick~\cite{sick03}), the Mainz low-$Q^2$
measurement (Bernauer {\it et al.}~\cite{bernauer10}) and this work compared
to the CODATA~\cite{mohr08} and muonic hydrogen spectroscopy (Pohl {\it et
al.}~\cite{pohl10}). The red dashed lines show the combined results from
CODATA, Bernauer {\it et al.} and this work while the black dotted lines show
the Pohl {\it et al.} uncertainty.}
\end{center}
\end{figure}

Several extractions of the proton charge radius are summarized in
Fig.~\ref{fig:radius}. Based on totally different data sets, the charge
radius extracted here is in excellent agreement with the Mainz extraction
$0.879\pm0.008$ fm~\cite{bernauer10} and the CODATA value
$0.8768\pm0.0069$ fm~\cite{mohr08}, which is mainly inferred from atomic
hydrogen spectroscopy. The combined result of the three measurements based
on electron--proton interactions gives:
\begin{equation}
\langle r^2_{E} \rangle^{1/2}_{ep} = 0.8772\pm0.0046~\mathrm{fm}.
\end{equation}
The Sick result is shown for comparison but not included in the average
because our updated extraction includes all of the data included in
Ref.~\cite{sick03}. A recent measurement of the Lamb shift in muonic
hydrogen~\cite{pohl10} yields a proton charge radius of $0.8418\pm0.0007$ fm,
which is 7.7$\sigma$ from the combined result of the $ep$ measurements.
Barring an error in the extraction of the muonic hydrogen result or a common
offset affecting both electron scattering and atomic hydrogen measurements,
there appears to be a true difference between electron-based and muon-based
measurements. This would seem to disfavor explanations that would lower the
CODATA result, e.g. based on a modified value of the Rydberg constant, and
suggests a missing correction that differs for electron and muon probes. One
mechanism that could cause such a difference was recently
proposed~\cite{miller11}. However, there is still significant activity aimed
at determining whether the discrepancy is the result of a missing element in
the QED corrections in the bound $\mu p$ system~\cite{jentschura11} or if a
more exotic explanation is required.

In conclusion, we present a new set of high precision form factor ratios
extracted using the recoil polarization technique.  These data provide improved
sensitivity to $G_M$ at low $Q^2$ while minimizing two-photon exchange
corrections which are of particular importance in separating $G_E$ and $G_M$.
Our results have been combined with previous cross section and polarization
measurements to provide an improved global analysis of the form factors. The
new data shift the extracted values of $G_E$ and $G_M$ at low $Q^2$, which
will impact the proton structure corrections to atomic energy levels in
hydrogen and muonic hydrogen as well as the extraction of the strange quark
contribution to the form factors probed in parity-violation $ep$
scattering~\cite{young06,liu07}.  While the updated form factors yield only a
small modification to the expected parity-violating asymmetry, typically less
than half of the assumed uncertainty, it will have a correlated effect on the
analysis of all such measurements.  Finally, we present an improved extraction of
the proton charge and magnetization radii, and find consistency between
electron-based probes of the proton radius, which disagree with the muonic
hydrogen result.

We thank the Jefferson Lab Physics and Accelerator Divisions. This work was
supported by the National Science Foundation, the Department of Energy,
including Contract No. DE-AC02-06CH11357, and the US-Israeli Bi-National
Scientific Foundation. Jefferson Science Associates operates the Thomas
Jefferson National Accelerator Facility under DOE Contract No.
DE-AC05-06OR23177.

\bibliography{ff_plb}

\begin{thebibliography}{42}
\expandafter\ifx\csname natexlab\endcsname\relax\def\natexlab#1{#1}\fi
\providecommand{\bibinfo}[2]{#2}
\ifx\xfnm\relax \def\xfnm[#1]{\unskip,\space#1}\fi
\bibitem[{Arrington et~al.(2007)Arrington, Roberts, and Zanotti}]{arrington07a}
\bibinfo{author}{J.~Arrington}, \bibinfo{author}{C.~D. Roberts},
  \bibinfo{author}{J.~M. Zanotti}, \bibinfo{journal}{J. Phys.}
  \bibinfo{volume}{G34} (\bibinfo{year}{2007}) \bibinfo{pages}{S23}.
\bibitem[{Perdrisat et~al.(2007)Perdrisat, Punjabi, and
  Vanderhaeghen}]{perdrisat07}
\bibinfo{author}{C.~F. Perdrisat}, \bibinfo{author}{V.~Punjabi},
  \bibinfo{author}{M.~Vanderhaeghen}, \bibinfo{journal}{Prog. Part. Nucl.
  Phys.} \bibinfo{volume}{59} (\bibinfo{year}{2007}) \bibinfo{pages}{694}.
\bibitem[{Arrington et~al.(2011)Arrington, de~Jager, and
  Perdrisat}]{arrington11a}
\bibinfo{author}{J.~Arrington}, \bibinfo{author}{K.~de~Jager},
  \bibinfo{author}{C.~F. Perdrisat}, \bibinfo{journal}{J.Phys.Conf.Ser.}
  \bibinfo{volume}{299} (\bibinfo{year}{2011}) \bibinfo{pages}{012002}.
\bibitem[{Gayou et~al.(2002)}]{gayou02}
\bibinfo{author}{O.~Gayou}, et~al., \bibinfo{journal}{Phys. Rev. Lett.}
  \bibinfo{volume}{88} (\bibinfo{year}{2002}) \bibinfo{pages}{092301}.
\bibitem[{Punjabi et~al.(2005)}]{punjabi05}
\bibinfo{author}{V.~Punjabi}, et~al., \bibinfo{journal}{Phys. Rev. C}
  \bibinfo{volume}{71} (\bibinfo{year}{2005}) \bibinfo{pages}{055202}.
  \bibinfo{note}{{Erratum-ibid. C {\bf 71}, (2005) 069902}}.
\bibitem[{Puckett et~al.(2010)}]{puckett10}
\bibinfo{author}{A.~J.~R. Puckett}, et~al., \bibinfo{journal}{Phys. Rev. Lett.}
  \bibinfo{volume}{104} (\bibinfo{year}{2010}) \bibinfo{pages}{242301}.
\bibitem[{Qattan et~al.(2005)}]{qattan05}
\bibinfo{author}{I.~A. Qattan}, et~al., \bibinfo{journal}{Phys. Rev. Lett.}
  \bibinfo{volume}{94} (\bibinfo{year}{2005}) \bibinfo{pages}{142301}.
\bibitem[{Guichon and Vanderhaeghen(2003)}]{guichon03}
\bibinfo{author}{P.~A.~M. Guichon}, \bibinfo{author}{M.~Vanderhaeghen},
  \bibinfo{journal}{Phys. Rev. Lett.} \bibinfo{volume}{91}
  (\bibinfo{year}{2003}) \bibinfo{pages}{142303}.
\bibitem[{Blunden et~al.(2003)Blunden, Melnitchouk, and Tjon}]{blunden03}
\bibinfo{author}{P.~G. Blunden}, \bibinfo{author}{W.~Melnitchouk},
  \bibinfo{author}{J.~A. Tjon}, \bibinfo{journal}{Phys. Rev. Lett.}
  \bibinfo{volume}{91} (\bibinfo{year}{2003}) \bibinfo{pages}{142304}.
\bibitem[{Chen et~al.(2004)Chen, Afanasev, Brodsky, Carlson, and
  Vanderhaeghen}]{chen04}
\bibinfo{author}{Y.~C. Chen}, \bibinfo{author}{A.~Afanasev},
  \bibinfo{author}{S.~J. Brodsky}, \bibinfo{author}{C.~E. Carlson},
  \bibinfo{author}{M.~Vanderhaeghen}, \bibinfo{journal}{Phys. Rev. Lett.}
  \bibinfo{volume}{93} (\bibinfo{year}{2004}) \bibinfo{pages}{122301}.
\bibitem[{Arrington et~al.(2011)Arrington, Blunden, and
  Melnitchouk}]{arrington11b}
\bibinfo{author}{J.~Arrington}, \bibinfo{author}{P.~G. Blunden},
  \bibinfo{author}{W.~Melnitchouk}  (\bibinfo{year}{2011}).
  \bibinfo{note}{ArXiv:1105.0951}.
\bibitem[{Friedrich and Walcher(2003)}]{friedrich03}
\bibinfo{author}{J.~Friedrich}, \bibinfo{author}{T.~Walcher},
  \bibinfo{journal}{Eur. Phys. J.} \bibinfo{volume}{A17} (\bibinfo{year}{2003})
  \bibinfo{pages}{607}.
\bibitem[{Crawford et~al.(2007)}]{crawford07}
\bibinfo{author}{C.~B. Crawford}, et~al., \bibinfo{journal}{Phys. Rev. Lett.}
  \bibinfo{volume}{98} (\bibinfo{year}{2007}) \bibinfo{pages}{052301}.
\bibitem[{Ron et~al.(2007)}]{ron07}
\bibinfo{author}{G.~Ron}, et~al., \bibinfo{journal}{Phys. Rev. Lett.}
  \bibinfo{volume}{99} (\bibinfo{year}{2007}) \bibinfo{pages}{202002}.
\bibitem[{Alcorn et~al.(2004)}]{alcorn04}
\bibinfo{author}{J.~Alcorn}, et~al., \bibinfo{journal}{Nucl. Instrum. Meth.}
  \bibinfo{volume}{A522} (\bibinfo{year}{2004}) \bibinfo{pages}{294}.
\bibitem[{Besset et~al.(1979)}]{besset79a}
\bibinfo{author}{D.~Besset}, et~al., \bibinfo{journal}{Nucl. Instrum. Meth.}
  \bibinfo{volume}{166} (\bibinfo{year}{1979}) \bibinfo{pages}{515}.
\bibitem[{Makino and Berz(1999)}]{cosy}
\bibinfo{author}{K.~Makino}, \bibinfo{author}{B.~Berz}, \bibinfo{journal}{Nucl.
  Instrum. Meth.} \bibinfo{volume}{A654} (\bibinfo{year}{1999}).
\bibitem[{Zhan(2010)}]{xiaohui_thesis}
\bibinfo{author}{X.~Zhan}, Ph.D. thesis, Massachusettes Institute of
  Technology, \bibinfo{year}{2010}. \bibinfo{note}{ArXiv:1108.4441}.
\bibitem[{Glister et~al.(2009)}]{glister09}
\bibinfo{author}{J.~Glister}, et~al., \bibinfo{journal}{Nucl. Instrum. Meth.}
  \bibinfo{volume}{A606} (\bibinfo{year}{2009}) \bibinfo{pages}{578}.
\bibitem[{Ron et~al.(2011)}]{ron11}
\bibinfo{author}{G.~Ron}, et~al.  (\bibinfo{year}{2011}).
  \bibinfo{note}{ArXiv:1103.5784}.
\bibitem[{Paolone et~al.(2010)}]{paolone10}
\bibinfo{author}{M.~Paolone}, et~al., \bibinfo{journal}{Phys. Rev. Lett.}
  \bibinfo{volume}{105} (\bibinfo{year}{2010}) \bibinfo{pages}{072001}.
\bibitem[{Milbrath et~al.(1998)}]{milbrath98}
\bibinfo{author}{B.~D. Milbrath}, et~al., \bibinfo{journal}{Phys. Rev. Lett.}
  \bibinfo{volume}{80} (\bibinfo{year}{1998}) \bibinfo{pages}{452}.
\bibitem[{Pospischil et~al.(2001)}]{pospischil01}
\bibinfo{author}{T.~Pospischil}, et~al., \bibinfo{journal}{Eur. Phys. J.}
  \bibinfo{volume}{A12} (\bibinfo{year}{2001}) \bibinfo{pages}{125}.
\bibitem[{Dieterich et~al.(2001)}]{dieterich01}
\bibinfo{author}{S.~Dieterich}, et~al., \bibinfo{journal}{Phys. Lett.}
  \bibinfo{volume}{B500} (\bibinfo{year}{2001}) \bibinfo{pages}{47}.
\bibitem[{Gayou et~al.(2001)}]{gayou01}
\bibinfo{author}{O.~Gayou}, et~al., \bibinfo{journal}{Phys. Rev. C}
  \bibinfo{volume}{64} (\bibinfo{year}{2001}) \bibinfo{pages}{038202}.
\bibitem[{Miller and Frank(2002)}]{miller02a}
\bibinfo{author}{G.~A. Miller}, \bibinfo{author}{M.~R. Frank},
  \bibinfo{journal}{Phys. Rev. C} \bibinfo{volume}{65} (\bibinfo{year}{2002})
  \bibinfo{pages}{065205}.
\bibitem[{Boffi et~al.(2002)}]{boffi02}
\bibinfo{author}{S.~Boffi}, et~al., \bibinfo{journal}{Eur. Phys. J.}
  \bibinfo{volume}{A14} (\bibinfo{year}{2002}) \bibinfo{pages}{17}.
\bibitem[{Faessler et~al.(2006)Faessler, Gutsche, Lyubovitskij, and
  Pumsa-ard}]{faessler06}
\bibinfo{author}{A.~Faessler}, \bibinfo{author}{T.~Gutsche},
  \bibinfo{author}{V.~E. Lyubovitskij}, \bibinfo{author}{K.~Pumsa-ard},
  \bibinfo{journal}{Phys. Rev. D} \bibinfo{volume}{73} (\bibinfo{year}{2006})
  \bibinfo{pages}{114021}.
\bibitem[{Belushkin et~al.(2007)Belushkin, Hammer, and Meissner}]{belushkin07}
\bibinfo{author}{M.~A. Belushkin}, \bibinfo{author}{H.~W. Hammer},
  \bibinfo{author}{U.~G. Meissner}, \bibinfo{journal}{Phys. Rev. C}
  \bibinfo{volume}{75} (\bibinfo{year}{2007}) \bibinfo{pages}{035202}.
\bibitem[{de~Melo et~al.(2009)de~Melo, Frederico, Pace, Pisano, and
  Salme}]{melo09}
\bibinfo{author}{J.~P. B.~C. de~Melo}, \bibinfo{author}{T.~Frederico},
  \bibinfo{author}{E.~Pace}, \bibinfo{author}{S.~Pisano},
  \bibinfo{author}{G.~Salme}, \bibinfo{journal}{Phys. Lett.}
  \bibinfo{volume}{B671} (\bibinfo{year}{2009}) \bibinfo{pages}{153}.
\bibitem[{Arrington et~al.(2007)Arrington, Melnitchouk, and
  Tjon}]{arrington07c}
\bibinfo{author}{J.~Arrington}, \bibinfo{author}{W.~Melnitchouk},
  \bibinfo{author}{J.~A. Tjon}, \bibinfo{journal}{Phys. Rev.}
  \bibinfo{volume}{C76} (\bibinfo{year}{2007}) \bibinfo{pages}{035205}.
\bibitem[{Bernauer et~al.(2010)}]{bernauer10}
\bibinfo{author}{J.~C. Bernauer}, et~al., \bibinfo{journal}{Phys. Rev. Lett.}
  \bibinfo{volume}{105} (\bibinfo{year}{2010}) \bibinfo{pages}{242001}.
\bibitem[{Sick(2003)}]{sick03}
\bibinfo{author}{I.~Sick}, \bibinfo{journal}{Phys. Lett. B}
  \bibinfo{volume}{576} (\bibinfo{year}{2003}) \bibinfo{pages}{62}.
\bibitem[{McKinley and Feshbach(1948)}]{mckinley48}
\bibinfo{author}{W.~A. McKinley}, \bibinfo{author}{H.~Feshbach},
  \bibinfo{journal}{Phys. Rev.} \bibinfo{volume}{74} (\bibinfo{year}{1948})
  \bibinfo{pages}{1759}.
\bibitem[{Arrington(2011)}]{arrington11c}
\bibinfo{author}{J.~Arrington}  (\bibinfo{year}{2011}).
  \bibinfo{note}{ArXiv:1008.3058, to appear in Phys.Rev.Lett.}
\bibitem[{Bernauer et~al.(2011)}]{bernauer11}
\bibinfo{author}{J.~C. Bernauer}, et~al.  (\bibinfo{year}{2011}).
  \bibinfo{note}{ArXiv:1108.3533}.
\bibitem[{Mohr et~al.(2008)Mohr, Taylor, and Newell}]{mohr08}
\bibinfo{author}{P.~J. Mohr}, \bibinfo{author}{B.~N. Taylor},
  \bibinfo{author}{D.~B. Newell}, \bibinfo{journal}{Rev. Mod. Phys.}
  \bibinfo{volume}{80} (\bibinfo{year}{2008}) \bibinfo{pages}{633}.
\bibitem[{Pohl et~al.(2010)}]{pohl10}
\bibinfo{author}{R.~Pohl}, et~al., \bibinfo{journal}{Nature}
  \bibinfo{volume}{466} (\bibinfo{year}{2010}) \bibinfo{pages}{213}.
\bibitem[{Miller et~al.(2011)Miller, Thomas, Carroll, and Rafelski}]{miller11}
\bibinfo{author}{G.~Miller}, \bibinfo{author}{A.~Thomas},
  \bibinfo{author}{J.~Carroll}, \bibinfo{author}{J.~Rafelski}
  (\bibinfo{year}{2011}). \bibinfo{note}{ArXiv:1101.4073}.
\bibitem[{Jentschura(2011)}]{jentschura11}
\bibinfo{author}{U.~Jentschura}, \bibinfo{journal}{Eur.Phys.J.}
  \bibinfo{volume}{D61} (\bibinfo{year}{2011}) \bibinfo{pages}{7}.
\bibitem[{Young et~al.(2006)Young, Roche, Carlini, and Thomas}]{young06}
\bibinfo{author}{R.~D. Young}, \bibinfo{author}{J.~Roche},
  \bibinfo{author}{R.~D. Carlini}, \bibinfo{author}{A.~W. Thomas},
  \bibinfo{journal}{Phys.Rev.Lett.} \bibinfo{volume}{97} (\bibinfo{year}{2006})
  \bibinfo{pages}{102002}.
\bibitem[{Liu et~al.(2007)Liu, McKeown, and Ramsey-Musolf}]{liu07}
\bibinfo{author}{J.~Liu}, \bibinfo{author}{R.~D. McKeown},
  \bibinfo{author}{M.~J. Ramsey-Musolf}, \bibinfo{journal}{Phys. Rev.}
  \bibinfo{volume}{C76} (\bibinfo{year}{2007}) \bibinfo{pages}{025202}.

\end{thebibliography}

\end{document}